# Solar-like oscillations in $\eta$ Boo

Hans Kjeldsen

*Institute of Physics and Astronomy, Aarhus University, DK-8000 Aarhus C, Denmark*

*and*

*European Southern Observatory, Karl-Schwarzschild-Str. 2, D-85748 Garching bei München, Germany*

Timothy R. Bedding[1]

*European Southern Observatory, Karl-Schwarzschild-Str. 2, D-85748 Garching bei München, Germany*

Michael Viskum and Søren Frandsen

*Institute of Physics and Astronomy, Aarhus University, DK-8000 Aarhus C, Denmark*

## ABSTRACT

We have observed evidence for $p$-mode oscillations in the G0 IV star $\eta$ Boo ($V = 2.68$). This represents the first clear evidence of solar-like oscillations in a star other than the Sun. We used a new technique which measures fluctuations in the temperature of the star via their effect on the equivalent widths of the Balmer lines. The observations were obtained over six nights with the 2.5 m Nordic Optical Telescope on La Palma and consist of 12684 low-dispersion spectra. In the power spectrum of the equivalent-width measurements, we find an excess of power at frequencies around $850\,\mu$Hz (period 20 minutes) which consists of a regular series of peaks with a spacing of $\Delta\nu = 40.3\,\mu$Hz. We identify thirteen oscillation modes, with frequency separations in agreement with theoretical expectations. Similar observations of the daytime sky show the five-minute solar oscillations at the expected frequencies.

astro-ph/9411016  3 Nov 94

---

[1]Current address: School of Physics, University of Sydney 2006, Australia. Electronic mail: bedding@physics.su.oz.au

1. INTRODUCTION

The remarkable success of helioseismology in probing the interior of the Sun has motivated many attempts to detect analogues of the solar five-minute oscillations in other stars. The measurement of oscillation frequencies would place important constraints on stellar model parameters and provide a strong test of evolutionary theory. However, despite several claims in the literature, it is fair to say that there has been no unambiguous detection of solar-like oscillations in any star except the Sun.

The difficulty with observing stellar oscillations lies in their very small amplitudes. Most previous attempts have used high-resolution spectroscopy to look for periodic Doppler shifts in spectral features (Gelly et al. 1986; Brown & Gilliland 1990; Innis et al. 1991; Brown et al. 1991; Pottasch et al. 1992; Edmonds 1993). A second method is differential CCD photometry, which measures fluctuations in the integrated stellar luminosity (Gilliland et al. 1993).

The observations reported here use a new technique which, like differential CCD photometry, exploits the fact that stellar oscillations are acoustic waves and hence affect the temperature of the star. We measure these temperature fluctuations, not via changes in the stellar luminosity, but rather through their effect on the equivalent widths of the Balmer lines.

The equivalent-width method is complementary to differential photometry, since the latter is applied to stellar clusters while the new method is used to observe isolated bright stars. However, unlike photometry, the measurement of equivalent widths is insensitive to atmospheric scintillation, because one is measuring the strength of a line relative to that of the continuum. Indeed, from several hours of test observations of $\alpha$ Cen taken with the ESO New Technology Telescope in July 1993, we have established that measuring the equivalent widths of broad absorption lines is limited only by photon noise. Furthermore, unlike the case of radial velocity measurements, moderate spectral resolution is sufficient. This allows a substantial improvement in signal-to-noise and removes the requirement for extreme instrumental stability.

2. METHOD

The Balmer series arises from transitions from the second to higher levels, which means that a hydrogen atom must first be excited before it can absorb a Balmer-line photon. Therefore, the equivalent width $W$ depends on the number of hydrogen atoms in the second level, which increases rapidly with temperature. The slope of this relation, as determined from model atmosphere calculations, is $\partial \log W/\partial \log T \simeq 6$ (Kurucz 1979; Gray 1992). This is valid for the H$\beta$, H$\gamma$ and H$\delta$ lines in G and F-type stars, and gives

$$\delta W/W \simeq 6 \, \delta T/T. \qquad (1)$$

What value of $\delta W/W$ can we expect to observe from stellar oscillations? The question of oscillation amplitudes has recently been addressed in detail by Kjeldsen & Bedding (1995; hereafter KB). In the Sun, the amplitudes of individual oscillation modes can vary considerably over timescales of several days. However, different modes vary independently and the maximum amplitude, taken over all modes, stays roughly constant. When the solar oscillations are measured in bolometric luminosity, this amplitude is $(\delta L/L)_{\rm bol} = 4.1\,\rm ppm$ (parts-per-million; see KB and references therein). This luminosity variation is due almost entirely to changes in temperature (the change in radius is negligible). Therefore, given that $L \propto R^2 T^4$, we can write $(\delta L/L)_{\rm bol} = 4\,\delta T/T$. From this we deduce that the oscillations in the Sun produce temperature fluctuations of about 1 ppm (6 milli-Kelvin). Equation 1 then implies an amplitude in Balmer-line equivalent width of 6 ppm.

Theoretical models predict that subgiants should have somewhat larger oscillation amplitudes than the Sun (Christensen-Dalsgaard & Frandsen 1983; Houdek et al. 1994). For this reason we have chosen to observe the star $\eta$ Boo, which is the brightest G-type subgiant in the sky (G0 IV, $V = 2.68$).

3. OBSERVATIONS AND REDUCTIONS

The observations of $\eta$ Boo were made over six nights (22–28 April 1994) using the 2.5 m Nordic Optical Telescope on La Palma. The data consist of 12684 grism spectra of the region 380–500 nm, at a dispersion of 0.2 nm per pixel. They were obtained using the Low Dispersion Spectrograph (secondary channel) and a 512 Tektronix CCD, with exposure times of 5 s and a dead time between exposures of 9 s. During the daylight hours we obtained spectra of the Sun by observing scattered light from the sky, with the aim of detecting the solar oscillations and confirming the validity of the equivalent-width method.

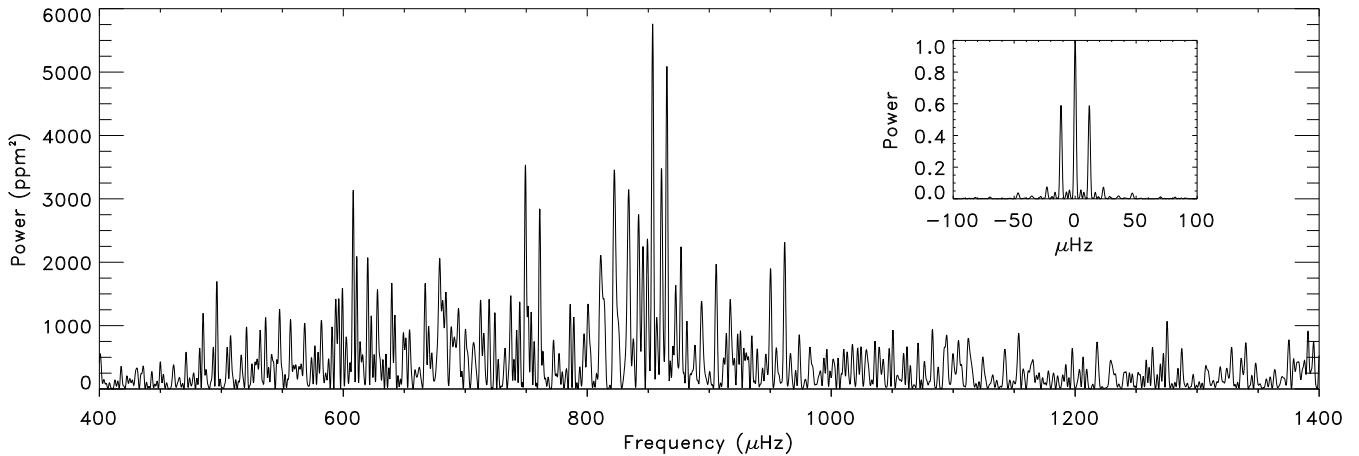

Fig. 1.— Power spectrum of equivalent width measurements for $\eta$ Boo. The inset shows the window function.

Processing of each spectrum consisted of the following steps: basic CCD calibration, including correction for nonlinearities at high intensity; extraction of a one-dimensional spectrum; and determination of the equivalent widths of the H$\beta$, H$\gamma$ and H$\delta$ lines by fitting their profiles. The resulting time series was decorrelated against external parameters to improve the estimates of equivalent-width measurements (Brown et al. 1991; Gilliland et al. 1993). The most important parameters for this procedure turned out to be the shape of the line, the structure of the continuum, the total intensity and the seeing. This decorrelation process reduced the rms scatter in the time series by a factor of 1.9. For the H$\beta$ data, which has the highest photon flux and the lowest noise, the rms of the final time series was 1030 ppm.

Finally, a discrete Fourier transform was used to calculate the power spectrum (PS), with each data point weighted according to the local rms scatter. This is shown in Figure 1. Note that the data have been high-pass filtered, which was necessary to carry out the decorrelation process referred to above. The filtering was performed by convolving the time series with a box-like function having FWHM 2200 s and subtracting this smoothed time series from the original. This filter, whose transfer function is shown in Figure 2, has little effect at frequencies above $\sim 700\,\mu$Hz.

In Figure 1 there is a clear hump of excess power centred at $850\,\mu$Hz, which we propose is due to oscillations in $\eta$ Boo. The corresponding period (20 min) is longer than for the Sun, which is expected for a star with a lower density and surface gravity (Brown 1991).

The background noise level in the rest of the PS is well fitted by a two-component model (KB): (i) white noise with an amplitude of 12.8 ppm and (ii) $1/f$ noise with an amplitude at $850\,\mu$Hz of 7.9 ppm. The contribution of photon noise is 12.2 ppm, indicating that the white-noise component comes almost entirely from photon statistics. The total number of photons collected from all three Balmer lines over the six nights was about $10^{10}$ photons Å$^{-1}$. The window function of the data, which is the PS of the sampling function, is shown in the inset. The daily gaps in the observations produce strong sidelobes at splittings of $\pm 1/\mathrm{day} = 11.57\,\mu$Hz.

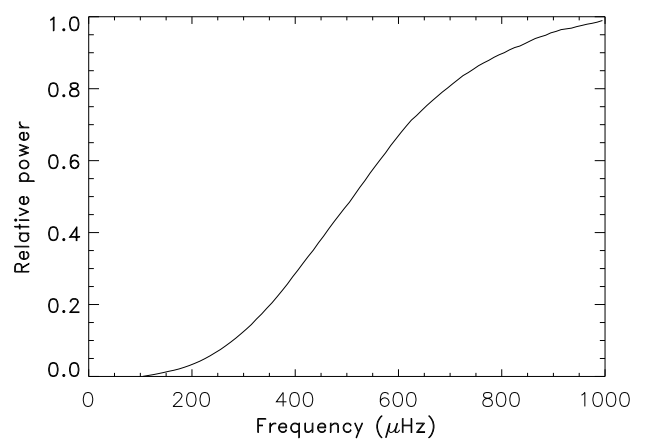

Fig. 2.— Transfer function of the high-pass filter applied to the $\eta$ Boo time series.

## 4. DISCUSSION

### 4.1. Search for a Comb-like Pattern

Confirmation of solar-like oscillations requires that we identify a series of peaks in the power spectrum similar to those seen for the Sun. For the solar oscillations, the strongest peaks are approximately equally spaced, corresponding alternately to modes with angular degrees $l = 0$ and $l = 1$. It is conventional to define the *large separation* $\Delta\nu$ as the frequency difference between modes that have the same value of $l$ and differ by one in radial order $n$. With this definition, the strongest peaks in the solar PS are spaced at intervals of $\frac{1}{2}\Delta\nu$.

One technique commonly used to search for periodicity in a PS is to calculate the power spectrum of the power spectrum (PS⊗PS). In the case of single-site observations, however, interpretation of the PS⊗PS is severely hampered by the daily gaps in the time series. For $\eta$ Boo, the PS⊗PS for the frequency range 750–950 $\mu$Hz shows evidence for periodicity at a frequency splitting of 40.3 $\mu$Hz. Simulations using an artificial oscillation signal plus noise indicate that the observed PS⊗PS is consistent with what is expected from genuine oscillations, although interpretation is difficult because of the gaps in the time series. To overcome this problem, we have developed a new method of searching for regularity in the PS.

We calculate a function, which we call the *comb response* (CR), as follows. Let $\nu_{\max}$ be the frequency of the strongest peak in the PS, which we tentatively identify as an oscillation mode. For the $\eta$ Boo data, $\nu_{\max} = 854\,\mu$Hz. Then, for each trial value of $\Delta\nu$, we calculate the product

$$\begin{aligned} CR(\Delta\nu) = & \ PS(\nu_{\max} - \tfrac{1}{2}\Delta\nu)\ PS(\nu_{\max} + \tfrac{1}{2}\Delta\nu) \\ & PS(\nu_{\max} - \Delta\nu)\ PS(\nu_{\max} + \Delta\nu) \\ & \left[PS(\nu_{\max} - \tfrac{3}{2}\Delta\nu)\ PS(\nu_{\max} + \tfrac{3}{2}\Delta\nu) \right. \\ & \left. PS(\nu_{\max} - 2\Delta\nu)\ PS(\nu_{\max} + 2\Delta\nu)\right]^{0.5}. \end{aligned} \quad (2)$$

A peak in the CR at a particular value of $\Delta\nu$ indicates the presence of a regular series of peaks in the PS, centred at $\nu_{\max}$ and having a spacing of $\frac{1}{2}\Delta\nu$.

A few comments about this choice of functional form are in order. It differs from a correlation function in that we take the product of the individual terms, rather than their sum. This means the CR is less likely to register a spurious response based on only a few strong peaks; if any peaks are missing, the CR will be reduced. We expect that a solar-like series of peaks should be modulated in amplitude by a broad envelope, and this is recognized by taking the square root of power at the outermost four frequencies. The function given above turned out to be good at correctly locating the regularity in simulated power spectra that contain artificial signal plus noise. An important advantage of the comb response over PS⊗PS is that the confusing effects of the window function are avoided. In the comb response, regularity in the PS appears quite naturally as a single strong peak.

Since the spacings between oscillation frequencies in the Sun are not exactly regular, and we expect a similar effect for $\eta$ Boo, we first smoothed the $\eta$ Boo PS to slightly lower resolution (2.2 $\mu$Hz FWHM). The comb response of this smoothed PS is shown in Figure 3. We indeed see a single peak at $\Delta\nu = 40.3\,\mu$Hz, consistent with the earlier PS⊗PS analysis.

### 4.2. Significance of the Detection

The evidence for solar-like oscillations of $\eta$ Boo consists of a hump of excess power in the PS, and an indication from the comb response of regularly-spaced peaks within that hump. How significant is this evidence?

We first ask whether one should be able to detect genuine oscillations in our data, given the noise level and sampling function. We have generated simulated time series consisting of artificial signal plus noise, each having exactly the same sampling function as the observations. The injected signal contained a solar-like frequency spectrum with a large separation of $\Delta\nu = 40.3\,\mu\text{Hz}$. The mode amplitudes were chosen to have a broad solar-like envelope. In each simulation, the phases of the oscillation modes were chosen at random and the amplitudes were randomized about their average values. All these characteristics were chosen to imitate as closely as possible the stochastic nature of oscillations in the Sun. Before calculating the power spectrum, we added normally-distributed noise to each time series, so as to produce a noise level in the PS of 15 ppm (consistent with the observed data).

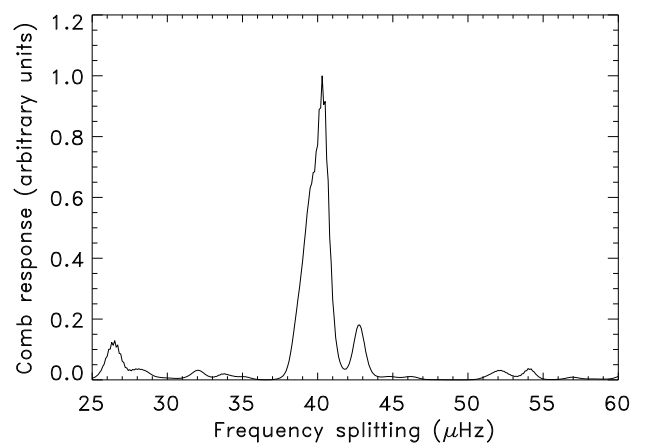

Fig. 3.— Comb response of the power spectrum of $\eta$ Boo data.

For simulations with signal having an average maximum amplitude of about 45 ppm, power spectra and comb responses very similar to those in Figures 1 and 3 occur often. This demonstrates that it is possible, in data like ours, to detect an oscillation spectrum with the characteristics suggested here. In short, we see exactly what one would expect to see from genuine oscillations.

A more difficult question is whether the observed features could be due solely to noise. Noise sources are usually either white (i.e., flat), follow a $1/f$ power law, or consist of single frequency. None of these could produce the observed hump of excess power in the PS of $\eta$ Boo. For this reason, it is difficult to conduct simulations to assess the performance of the comb-response analysis. Producing a hump of excess power simply by enhancing a flat noise spectrum with a Gaussian envelope failed to reproduce the observed distribution of peak heights. A closer match to the $\eta$ Boo PS could be made by injecting sinusoidal signals, in the same way as above, but with randomized frequencies. From these simulations, we could assess how likely the comb response was to produce a peak, even when there is no deliberate regularity in the power spectrum. The result is that the CR sometimes contains peaks greater than that seen in the $\eta$ Boo data (about 20–30% of the time). On the basis of this, there is a possibility that the observed regularity is a chance effect, but only if we assume an unrealistic noise model consisting of power at individual frequencies.

In the analysis so far, we have been treating the six-night data set as a whole. We have also examined whether the signal is present in subsets of the data. When the time series is divided into two sets of three nights, the hump of excess power is present in both sets. The same is true when the data are taken two nights at a time, but to a lesser extent because of the lower signal-to-noise.

### 4.3. Identification of Oscillation Frequencies

Assuming the $40.3\,\mu\text{Hz}$ regularity to be genuine, we now attempt to identify the oscillation frequencies directly in the power spectrum. Once again, the strong sidelobes of the window function complicate matters and we address this problem using a modified version of the CLEAN algorithm (Roberts et al. 1987). The modification consists of using the measured regularity ($40.3\,\mu\text{Hz}$) to help prevent the algorithm from selecting sidelobes instead of central peaks. The following is a brief description of the procedure.

We first identify the strongest peak in the PS in the frequency range of interest (700–1000 $\mu$Hz) and subtract the corresponding sine wave from the time series. We then recompute the PS and define a new search area that is centred a distance $40.3\,\mu\text{Hz}$ from the first component. The radius of the search area is $1.5\,\mu\text{Hz}$. If this region contains a peak having a strength > 25 ppm, we identify this peak as the second CLEAN component, subtract the corresponding sine wave from the time series and recompute the PS. This process is repeated, moving in multiples of $40.3\,\mu\text{Hz}$, first in one direction and then the other. Once the whole region of interest (700–1000 $\mu$Hz) has been covered, the strongest remaining peak is identified and the process begins again with a new set of search areas separated by $40.3\,\mu\text{Hz}$. The procedure stops when all remaining peaks are below 25 ppm.

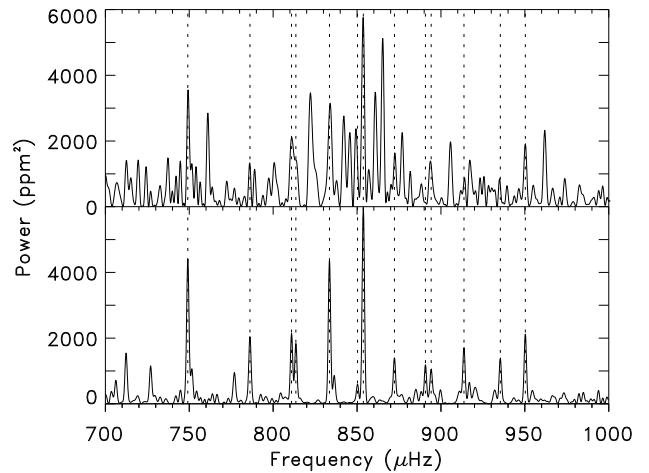

Fig. 4.— Power spectrum of $\eta$ Boo data (upper panel), and the power spectrum after CLEANing (lower panel). The dashed lines indicate CLEAN components that we identify as oscillation modes.

For improved accuracy, each time a new component is identified, the parameters of all previous CLEAN components are recalculated. That is, in each iteration we return to the original time series and fit the frequencies, phases and amplitudes of all components simultaneously. The components, having been slightly adjusted, are then subtracted from the original time series and the PS is calculated, ready for the next iteration.

The result of applying this algorithm to the $\eta$ Boo data is shown in Figure 4. Of twenty components found in the CLEANing process, we identify thirteen as possible oscillation modes (the rest are probably noise peaks). A useful way of displaying the mode frequencies is with an echelle diagram, and this is shown in Figure 5. This diagram is similar to that seen for the Sun (e.g., Toutain & Fröhlich 1992), leading us to assign $l$-values as shown. Three points depart significantly from the regular pattern, and may be modes that are shifted by the phenomenon of avoided crossings (Christensen-Dalsgaard et al. 1995).

The oscillation frequencies (in $\mu$Hz) are given in Table 1. Uncertainties in the frequencies range from $0.2\,\mu$Hz for the strongest peaks to about $1\,\mu$Hz for the weakest. The values of $n$ were based on the first-order relation $\nu_{n,l} \simeq (n + l/2 + \epsilon)\Delta\nu$ (Vandakurov 1968), which also gives $\epsilon = 1.18$.

The last row in Table 1 gives the large separation for each value of $l$, defined as $\Delta\nu_l \equiv \nu_{n+1,l} - \nu_{n,l}$. The uncertainties include measurement errors and the departure of the frequencies from strict regularity. The three discrepant modes, marked with asterisks in Table 1, were not used to calculate $\Delta\nu_1$.

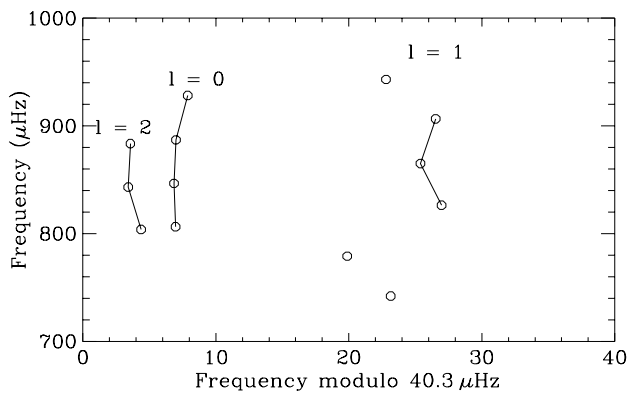

Fig. 5.— Echelle diagram of $\eta$ Boo oscillations.

### 4.4. Comparison with Theory

Comparison of oscillation frequencies and amplitudes with theory requires knowledge of the luminosity, temperature and mass of the star. According to Bell & Gustafsson (1989) and Blackwell & Lynas-Gray (1994), the effective temperature of $\eta$ Boo is $T_{\rm eff} = 6050 \pm 60$ K and its angular diameter is $2.24 \pm 0.02$ mas (milliarcsec). These estimates are based on the infrared flux method; a direct measurement using interferometry would be valuable. Using the parallax of $85.8 \pm 2.3$ mas (Harrington et al. 1993), it is straightforward to calculate the stellar radius and luminosity: $R = 2.80 \pm 0.08\,R_\odot$ and $L = 9.5 \pm 0.7\,L_\odot$.

Estimating the mass requires reference to evolutionary models. The metallicity is greater than solar: [Fe/H] $\simeq 0.19$ (Tomkin et al. 1985; Edvardsson et al. 1993). Based on published models of metal-rich stars (Schaerer et al. 1993; Fagotto et al. 1994; Bressan et al. 1993), we estimate a mass for $\eta$ Boo of $M = 1.6 \pm 0.2\,M_\odot$.

To a good approximation, the large separation should only depend on the mean density of the star. Scaling from the solar case implies $\Delta\nu \simeq 36 \pm 3\,\mu$Hz for $\eta$ Boo, in reasonable agreement with the observations. There are two small separations which reflect second-order departures from asymptotic theory. Their definitions and observed values are:

$$\delta\nu_{01} \equiv \tfrac{1}{2}(\nu_{n,0} + \nu_{n+1,0}) - \nu_{n,1}$$
$$= 0.9 \pm 0.5\,\mu\text{Hz}$$

and

$$\delta\nu_{02} \equiv \nu_{n+1,0} - \nu_{n,2}$$
$$= 3.1 \pm 0.3\,\mu\text{Hz}.$$

Since these separations arise from a second-order term

Table 1: Oscillation frequencies for $\eta$ Boo

|  | $l = 0$ | $l = 1$ | $l = 2$ |
| --- | --- | --- | --- |
| $n = 17$ |  | 749.1* |  |
| $n = 18$ |  | 786.2* | 811.0 |
| $n = 19$ | 813.6 | 833.6 | 850.3 |
| $n = 20$ | 853.7 | 872.3 | 890.8 |
| $n = 21$ | 894.2 | 913.7 |  |
| $n = 22$ | 935.4 | 950.3* |  |
| $\Delta\nu_l$ | $40.6 \pm 0.3$ | $40.1 \pm 0.7$ | $39.9 \pm 0.4$ |

* Not used in calculating $\Delta\nu_1$

proportional to $l(l + 1)$ (Tassoul 1980), we expect to find $\delta\nu_{02} \simeq 3\,\delta\nu_{01}$, which is indeed the case. The actual values of the small separations are sensitive to stellar age. Theoretical models by Christensen-Dalsgaard (1988) predict $\delta\nu_{02} = 2$–$4\,\mu$Hz for a star of this spectral type and luminosity class. Again, the agreement with observations is good. To make further progress, we are constructing detailed models of $\eta$ Boo and comparing the calculated oscillation frequencies with the values in Table 1 (Christensen-Dalsgaard et al. 1995).

We estimate the average amplitude of the oscillations, as defined in Section 2., to be $\sim 45$ ppm. This estimate was derived from the distribution of peak heights (after allowing for the noise level; see KB), and was confirmed in the simulations described above. Based on theoretical models, we expect the amplitudes of stellar oscillations to scale roughly as $L/(MT_{\mathrm{eff}})$ (KB, equations 5 and 6). For $\eta$ Boo, this implies an amplitude $\sim 5.3$ times greater than solar, or 35 ppm. Thus, the signal we have detected is approximately at the expected level. For comparison, we note that if the same oscillation signal (45 ppm) were observed in radial velocity, it would have an amplitude of $1.6\,\mathrm{m\,s^{-1}}$. Without a larger telescope, such a measurement would be very difficult in a star with the apparent magnitude of $\eta$ Boo (Brown et al. 1991).

### 4.5. Solar Data

During the daytime, we obtained spectra of the sky with the aim of detecting the solar oscillations. Their expected amplitude (6 ppm in equivalent width) is much lower than that of $\eta$ Boo, which makes detection difficult. A compensating factor is the larger number of photons available from long-slit observations of the sky.

The solar data were processed in the same manner as for $\eta$ Boo. The final power spectrum was photon-noise limited in the frequency range of the solar oscillations (around 3 mHz), with a noise level of 5.1 ppm. This is indeed lower than for $\eta$ Boo, thanks to the larger number of photons, but is still too high to allow a clear detection of the solar signal. In particular, we found no obvious excess in the power spectrum at frequencies near 3 mHz.

The comb-response method and similar techniques are not useful at this low level of signal-to-noise. Nevertheless, since the frequencies of the solar oscillations are well known, we can attempt to verify whether a signal is present in our data at the expected level. To do this, we have "folded" our solar power spectrum using the known large separation of the Sun ($\Delta\nu = 134.9\,\mu$Hz; Toutain & Fröhlich 1992). That is, we averaged together successive segments of length $\Delta\nu$ in the vicinity of 3 mHz. The result is shown in Figure 6. We identify four peaks — and their sidelobes — corresponding to degrees $l = 0$, 1, 2 and 3. Each of the four peaks in the figure coincides closely (within $1.5\,\mu$Hz) with the expected position.

To test the significance of this result, we have again made realistic simulations. In each simulation, we generated a time series containing signal at the known frequencies of the solar oscillations (Libbrecht et al. 1990), with random phases. The oscillation amplitudes were set according to the broad envelope that is seen in high S/N observations of the Sun. The noise level and sampling function of the time series were the same as for the actual observations. Ten separate simulations were generated, each using a different random number seed for the mode phases and the noise source.

For each simulated time series we calculated the power spectrum and the folded power spectrum, in the same way as for the real data. The simulated power spectra are indistinguishable from the observations: all show a flat noise level with little or no excess power around 3 mHz. On the other hand, all the folded power spectra show evidence of the injected signal, with varying degrees of significance. This vari-ation arises from the stochastic behaviour of the noise and the interaction of modes with different phases. Importantly, the positions of the peaks in the folded PS vary by a small amount (up to $2\,\mu$Hz) over the ensemble of ten simulations, consistent with the real solar data.

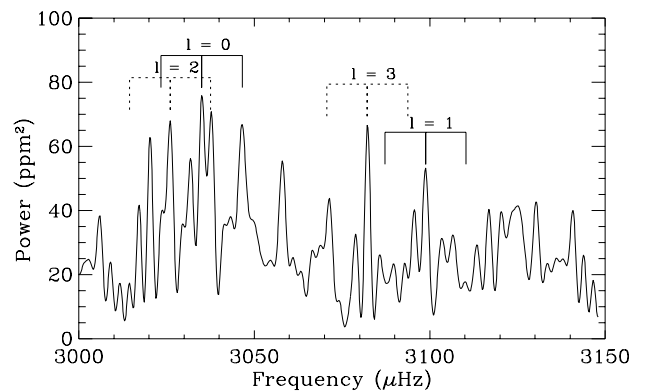

Fig. 6.— Folded power spectrum of solar observations. We identify four strong peaks corresponding to modes with degrees $l = 0$, 1, 2 and 3. For each, we also show the positions of the 1/day sidelobes ($\pm 11.57\,\mu$Hz).

Note that only two pieces of information about the solar oscillations have been used in producing the folded PS in Figure 6: the large separation ($134.9\,\mu$Hz) and the approximate frequency of the modes (around 3 mHz). We injected no information about the absolute frequencies, but are able to recover them within the uncertainties implied by the simulations. This is strong evidence that the solar oscillations are present in the data and that the equivalent-width method is valid.

## 5. CONCLUSIONS AND FUTURE PROSPECTS

We have presented evidence for solar-like oscillations in $\eta$ Boo that have amplitudes and frequency separations in good agreement with theoretical expectations. Although we believe this evidence is convincing, we recognize from the number of previous claims in the literature that caution is appropriate.

One puzzle is the apparent detection of the $l = 3$ modes in the Sun, which is surprising given that the observations did not resolve the solar disk. Perhaps the equivalent widths of Balmer lines contain some spatial information, through effects analogous to limb darkening or brightening. This clearly deserves study, since the ability to measure $l = 3$ modes in other stars would be a significant bonus. Unfortunately, we cannot do this with the present $\eta$ Boo data, because the expected position of the $l = 3$ modes is close to a 1/day sidelobe of $l = 0$. Future multi-site observations to provide continuous coverage would be highly desirable.

Turning to F-type subgiants, Procyon (F5 IV, $V = 0.4$) is an obvious target to observe with the equivalent-width method. Recent observations of the open cluster M67 suggest that oscillations in F stars must have amplitudes less than has generally been assumed (Gilliland et al. 1993; KB). A detection in Procyon would clearly be important. Assuming the same instrumental configuration and duty cycle as was used for $\eta$ Boo, one could reach a noise level in six nights on Procyon of 2.5–3 ppm, thanks to the fact that it is both brighter than $\eta$ Boo and has much stronger Balmer lines. Such observations would have a sensitivity 2–3 times better than has been previously achieved for that star (Brown et al. 1991; KB).

Observations of main sequence stars are more demanding because of the lower oscillation amplitudes. The best target is $\alpha$ Cen A (G2 V, $V = 0.0$), which should have an amplitude in equivalent width of $\sim 8$ ppm. As discussed in detail by Brown et al. (1994), our accurate knowledge of this star's fundamental parameters means that oscillation measurements would be particularly valuable for testing stellar evolutionary theory.

We would be happy to provide ASCII files of the time series and power spectra. Please direct inquiries to TRB at bedding@physics.su.oz.au.

We are grateful to Anton N. Sørensen and Bjarne Thomsen for excellent support with the observations, and to Tim Brown, Jørgen Christensen-Dalsgaard, Lawrence Cram, Peter Edmonds and Ron Gilliland for valuable discussions. We thank Albert Zijlstra and Dante Minniti for comments on the manuscript. HK acknowledges financial support from the Carlsberg Foundation and the Danish Board for Astronomical Research.

REFERENCES

Bell, R.A., & Gustafsson, B., 1989, MNRAS 236, 653

Blackwell, D.E., & Lynas-Gray, A.E., 1994, A&A 282, 899

Bressan, A., Fagotto, F., Bertelli, G., & Chiosi, C., 1993, A&AS 100, 647

Brown, T.M., 1991, The Sun and asteroseismology. In: Lambert, D.L. (ed.), Frontiers of Stellar Evolution, p. 139, ASP Conference Series, Vol. 20

Brown, T.M., & Gilliland, R.L., 1990, ApJ 350, 839

Brown, T.M., Gilliland, R.L., Noyes, R.W., & Ramsey, L.W., 1991, ApJ 368, 599

Brown, T.M., Christensen-Dalsgaard, J., Weibel-Mihalas, B., & Gilliland, R.L., 1994, ApJ 427, 1013

Christensen-Dalsgaard, J., 1988, A Hertzsprung-Russell diagram for stellar oscillations. In: Christensen-Dalsgaard, J., & Frandsen, S. (eds.), Proc. IAU Symp. 123, Advances in Helio- and Asteroseismology, p. 295, Kluwer, Dordrecht

Christensen-Dalsgaard, J., & Frandsen, S., 1983, Sol. Phys. 82, 469

Christensen-Dalsgaard, J., Bedding, T.R., & Kjeldsen, H., 1995, in preparation

Edmonds, P.D., 1993, *Asteroseismology,* PhD thesis, University of Sydney

Edvardsson, B., Andersen, J., Gustafsson, B., Lambert, D.L., Nissen, P.E., & Tomkin, J., 1993, A&A 275, 101

Fagotto, F., Bressan, A., Bertelli, G., & Chiosi, C., 1994, A&AS 104, 365

Gelly, G., Grec, G., & Fossat, E., 1986, A&A 164, 383

Gilliland, R.L., Brown, T.M., Kjeldsen, H., McCarthy, J.K., Peri, M.L., Belmonte, J.A., Vidal, I., Cram, L.E., Palmer, J., Frandsen, S., Parthasarathy, M., Petro, L., Schneider, H., Stetson, P.B., & Weiss, W.W., 1993, AJ 106, 2441

Gray, D.F., 1992, *The observation and analysis of stellar photospheres,* Vol. 20 of *Cambridge Astrophysics Series,* Cambridge: Cambridge University Press, 2nd edition

Harrington, R.S., Dahn, C.C., Kallarakal, V.V., Guetter, H.H., Riepe, B.Y., Walker, R.L., Pier, J.R., Vrba, F.J., Luginbuhl, C.B., Harris, H.C., & Ables, H.D., 1993, AJ 105, 1571

Houdek, G., Rogl, J., Balmforth, N.J., & Christensen-Dalsgaard, J., 1994, Excitation of solarlike oscillations in main–sequence stars. In: Ulrich, R.K., Rhodes, Jr, E.J., & Däppen, W. (eds.), *GONG '94: Helio- and Astero-seismology from Earth and Space,* ASP Conference Series, in press

Innis, J.L., Isaak, G.R., Speake, C.C., Brazier, R.I., & Williams, H.K., 1991, MNRAS 249, 643

Kjeldsen, H., & Bedding, T.R., 1995, A&A, in press (KB)

Kurucz, R.L., 1979, ApJS 40, 1

Libbrecht, K.G., Woodard, M.F., & Kaufman, J.M., 1990, ApJS 74, 1129

Pottasch, E.M., Butcher, H.R., & van Hoesel, F.H.J., 1992, A&A 264, 138

Roberts, D.H., Lehár, J., & Dreher, J.W., 1987, AJ 93, 968

Schaerer, D., Charbonnel, C., Meynet, G., Maeder, A., & Schaller, G., 1993, A&AS 102, 339

Tassoul, M., 1980, ApJS 43, 469

Tomkin, J., Lambert, D.L., & Balachandran, S., 1985, ApJ 290, 289

Toutain, T., & Fröhlich, C., 1992, A&A 257, 287

Vandakurov, Y.V., 1968, Sov. Astron. 11, 630